\documentclass[english,aps,twocolumn,showpacs]{revtex4}
\usepackage{graphicx}
\usepackage{amssymb}
\usepackage{babel}

\begin{document}

\title{Pseudo-nonstationarity in the scaling exponents of finite interval time series}

\author{K. H. Kiyani}

\email{k.kiyani@warwick.ac.uk}

\author{S. C. Chapman}

\affiliation{Centre for Fusion, Space and Astrophysics; Department of Physics,
University of Warwick, Gibbet Hill Road, Coventry, CV4 7AL, United
Kingdom}

\author{N. W. Watkins}

\affiliation{British Antarctic Survey, High Cross, Madingley Road, Cambridge CB3 0ET, United Kingdom}

\begin{abstract}
The accurate estimation of scaling exponents is central in the observational study of scale-invariant phenomena. Natural systems unavoidably provide observations over restricted intervals; consequently a stationary stochastic
process (time series) can yield anomalous time variation in the scaling exponents, suggestive of non-stationarity. The variance in the estimates of scaling exponents computed from an interval of $N$ observations is known for finite variance processes to vary as $\sim 1/N$ as $N \rightarrow \infty$ for certain statistical estimators; however, the convergence to this behaviour will depend on the details of the process, and may be slow. 
We study the variation in the scaling of second order moments of the time series increments with $N$, for a variety of synthetic and `real world' time series; and find that in particular for heavy tailed processes, for realizable $N$, one is far from this  $\sim 1/N$ limiting behaviour. We  propose a semi-empirical estimate for the minimum $N$ needed to make a meaningful estimate of the scaling exponents for model stochastic processes and compare these with some `real world' time series.
\end{abstract}


\maketitle

\section{Introduction}

Testing for, and quantifying scaling is central to the application of statistical theories to `real-world' extended systems. A broad range of theoretical frameworks such as turbulence \cite{Frisch1995}, critical phenomena \cite{Sethna2001} and complex systems \cite{Sornette2000} frame their predictions in terms of the statistical properties of (arbitrarily large) ensembles as a function of scale. Under the assumption of ergodicity the statistical scaling property of an extended system is captured to some extent by a reduced (embedded) set of observations or measurements; so that a 1-D cut through a $N$ dimensional system will be sufficient to indicate whether scaling is present, and in a quantitative way can usefully restrict the scaling exponents of the system as a whole. This approach is pragmatic -- in physical systems it is generally not practicable to capture and analyze the full spatiotemporal information of all points in the system on all scales. A key observable is then the quantitative scaling properties of such a one dimensional sample or time series. An example of this is the use of the Taylor hypothesis in turbulence, where the time series at a single point is used as a proxy for the statistical properties of the flow \cite{Pope2000}.

Time series are also often parsed into sub-intervals to isolate processes of interest, or to remove features which might contaminate the calculation of the quantity of interest. Examples of this in the study of solar wind turbulence are the separating of fast and slow wind, and open/closed field line regions \cite{Hnat2004,Smith2006}; isolating or removing signals of interplanetary shocks, magnetosheath crossings, and coronal mass ejection remnants \cite{Retino2007,Sundkvist2007}; or where the interval is restricted by a secular change in parameters as the spacecraft changes location \cite{Sorriso-valvo2007,nicol2008}.
Examples in the study of the earth's geomagnetic field include isolating `quasi-stationary' and `quiet' intervals in magnetic field data \cite{Smith2006,Jankovicova2008}; and the effects of finite sample size in the power spectral exponent estimates in the ionosphere by ground-based measurements \cite{abel2002}. `Locally stationary processes' are also discussed in speech signal analysis \cite{Mallat1999} and physiological non-stationary signals \cite{Popivanov1999}; and of course statistical forecasting, whether in the context of  seasonal weather or the financial markets \cite{Leontitsis2006}, is based on time series histories which rely on the stationarity assumption. In all of these cases, it is intuitively apparent that smaller data intervals will result in poorer statistics, which will be manifest in the variance of the observed values of the exponents. The observation of a (non-secular) variation in the scaling exponents therefore has two interpretations; either it is due to intrinsic fluctuations as a result of the finite $N$ interval, or it is a consequence of non-time stationarity of the time series $x(t)$ i.e. different scaling behaviour due to different physical phenomena. If the properties of the underlying process are not known \emph{a priori} we need a method to distinguish these two interpretations in a quantitative manner; or at best to put a degree of confidence that it is due to one and not the other. 

The most commonly used tool to both establish and quantify scaling in a time series $x(t)$ is to test for scaling of the power spectral density $F(\omega)\sim \omega^{-\beta}$, and obtain the exponent $\beta$. In a physical system, such scaling can only be observed over a finite range, limited by the interval (in time $t$)  of $N$ observations  over which the system is measured. 
From large-sample theory (asymptotic limit of $N\rightarrow \infty$) the variance in the power spectral exponent $\beta$ computed using least squares regression from an interval of $N$ samples is known \cite{robinson1995,Beran1994} for finite variance processes to vary as $\sim 1/N$ as $N \rightarrow \infty$. 
One method to obtain more complete information about the scaling properties of a stochastic process $x(t)$ is captured by how the statistical properties of the increments $y(t,\tau)=x(t+\tau)-x(t)$ vary with the differencing scale $\tau$; the differencing being a particular type of coarse-graining operation which has been chosen due to the easy analogy with random walks, return probabilities etc. However, there exist other coarse-graining operations which although more involved, possess additional highly desirable properties when studying scaling. In particular, wavelets which (with some wavelet functions) when combined with their detrending capabilities have been shown to be a natural and computationally efficient way of studying scale-by-scale statistical behaviour \citep{Muzy1993,abry2003,Mallat1999}.
In this paper we will discuss the behaviour of the scaling properties of the second order moment $\left\langle y(t,\tau)^2 \right\rangle\sim \tau^{\zeta(2)}$. We may anticipate that the statistical properties of this scaling exponent $\zeta(2)$ follow that of $\beta$; indeed there exist many results for a range of different estimators of the $\zeta(2)$ \cite{stoev2002,Bardet2000} that directly show the asymptotic $\sim 1/N$ behaviour discussed above.
In practice, the convergence to this $\sim 1/N$ behaviour will depend on the details of the process and the estimator and, as we shall show in this paper, is often slow.

An essential tool in the analysis of `real world' time series in the context of scaling is then a prescription for the variance in the scaling exponents of $x(t)$ as a function of the number of observations  $N$ in the chosen interval. In this paper we make some first steps in this direction by obtaining empirical estimates from the study of a variety of stochastic processes that have been used as models for physical systems. We focus on finite size $N$ realizations of self-affine cases with Gaussian distributed increments in the form of a standard Brownian motion and  fractional Brownian motion (fBm); and those with heavy tails, namely  $\alpha$-stable L\'evy motion and linear fractional stable motion (LFSM) \cite{samorodnitsky1994,stoev2004a,stoev2004b}. A representative case for multifractal scaling is provided by the $p$-model, often used to characterize observations of turbulence \cite{meneveau1987,meneveau1990}.

The fundamental property of ergodicity in systems that exhibit scaling implies time stationarity. In its strong sense time stationarity implies that the probability density function (PDF) of $x(t)$ does not change with time; this is known as strict stationarity. Pragmatically, weak stationarity, that is time independence of the variance or second order moment is usually adopted -- the latter convenience is usually assumed due to the special place that the Central Limit Theorem and the Gaussian distribution hold in statistics. 
In this paper we are concerned specifically with the behaviour of scaling exponents which are characterized through the statistical properties of the increments $y(t,\tau)$, rather than the time series $x(t)$; hence we will use as our test time series examples that have stationarity in $y(t,\tau)$, and not in $x(t)$. 
 
We will focus on the statistics of the scaling exponent of the second order moment of the increments, as this also captures the power spectral exponent $\beta$, and for self-affine finite-variance processes the Hurst exponent $H$ (see next section and also \cite{Kiyani2006} for the infinite variance case). We will study these processes for a range of values of $N$ that are feasable for realisable physical systems; and find that in particularly for the heavy-tailed processes, the variance in the exponent with $N$ shows a significant departure from the $1/N$ asymptotic behaviour. Nevertheless, for these heavy-tailed processes, we find empirical evidence of an intermediate range of scaling with $N^{-\gamma}$. We will estimate the time series interval $N$ required to capture the scaling exponent to reasonable precision; this places a lower limit on the sample size. 
A related study to this was conducted to investigate and compare the effects of finite sample size on different statistical estimators for the Hurst exponent $H$ for a Gaussian white noise process \cite{Weron2002}.
Stationarity also implies a particular PDF of the values of the exponent obtained from many, length $N$, realisations of a given process. This is known asymptotically for $N\rightarrow \infty$ for the processes based on Gaussian increments (generalizable to finite variance processes) and is also known in this asymptotic sense for infinite variance processes; both processes approaching a Gaussian distribution for the scaling exponents as $N\rightarrow \infty$  \cite{Beran1994,robinson1995,velasco1999,stoev2002,stoev2004b,EmbrechtsMaejima2002} (using least squares and maximum likelihood estimation schemes). For the intermediate stage of finite $N$ we find intermediate distributions for the exponents; resembling both the asymptotic Gaussian forms and, for heavy-tailed data, Gumbel max-stable (Extreme value type I) distributions.  Comparing these results with that found for real-world time series may provide an additional test for stationarity in the increments. In this spirit we finally illustrate these ideas with some examples of real-world time series in the form of \emph{in-situ} observations of magnetic field and velocity in the turbulent solar wind using data from spacecraft at 1AU in the ecliptic; and comment on the statistical properties of their scaling exponents in light of the representative synthetic toy models.

\section{Methodology}

We will focus attention on the scaling exponent $\zeta(2)$ of the second order moment of the increments also known as the second order structure function:
\begin{equation}
\left\langle y(t,\tau)^2 \right\rangle =\left\langle \left( x(t+\tau)-x(t) \right) ^2 \right\rangle = \left\langle y(t,1)^2 \right\rangle \tau^{\zeta(2)} \ ,
\label{eq:minus1}
\end{equation}
where we have assumed that the increment process is at least second order stationary i.e. $\left\langle y(t,\tau)^2 \right\rangle=\left\langle y(\tau)^2 \right\rangle$ (weak-stationarity). In particular, this implies that the power spectral density of a discrete time random walk $x(t)$ of \emph{i.i.d.} stationary increments with finite variance, scales as \cite{Mallat1999}
\begin{equation}
F(\omega) \sim \omega^{-(\zeta(2)+1)}\ ,\label{eq:0}
\end{equation}
where the scaling exponent $\zeta(2)$ is related to the power spectral exponent $\beta$ of $x(t)$ by $\zeta(2)=\beta-1$. For self-affine process with Hurst exponent $H$ the PDF $P(y,\tau)$ of the increments obeys the scaling relation (for the case of $\alpha$-stable processes with finite $N$ see \cite{Kiyani2006}, and the discussion to follow)
\begin{equation}
 P(y,\tau)=\tau^{-H}\mathcal{P}^{s}(\tau^{-H}y)\ , \label{eq:1}
\end{equation}
where the PDF $P$ at any scale $\tau$ can be collapsed onto a unique scaling function $\mathcal{P}^{s}$. The scaling relation (\ref{eq:1})
implies that the scaling of the structure functions to all orders $p$ \cite{Kiyani2006} is given by
$\left\langle y(\tau)^p \right\rangle = \left\langle y(1)^p \right\rangle \tau^{\zeta(p)}$ where
 $\zeta(p)=Hp$; and thus we have that $\zeta(2)=2H$. Our results concerning the statistical behaviour of $\zeta(2)$ with $N$ will thus also apply to the power spectral exponent $\beta$ for all the models concerned and the Hurst exponent $H$ for the self-affine models; both are commonly used to characterize statistical scaling. 
Our remarks can also be generalized to the scaling exponents $\zeta(p)$ of structure functions of higher-order positive moments. These are relevant for multifractal processes where the $\zeta(p)$ are a nonlinear function of $p$ and so $H$ or $\beta$ are not sufficient to determine the complete statistical scaling of the $y(t,\tau)$.

Our study consists of partitioning a given time series $x(t)$ into $L$ equal intervals of sample size $N$ denoted by $x_{i}(t)$ where $i=1\ldots L$. Each of these intervals are then differenced on scale $\tau$ to produce a time series of the increments $y_{i}(t,\tau)=x_{i}(t+\tau)-x_{i}(t)$ of the process $x_{i}(t)$.

We will look for scaling of the second order moment (structure function)
\begin{equation}
 M_{i}^{2}(\tau)=\left\langle y_{i}(\tau)^2 \right\rangle=\int^{y_i^+}_{y_i^-}y_{i}^{2}P_{i}(y_{i},\tau)dy_{i}\ , \label{eq:2}
\end{equation}
with $\tau$ such that  $M_{i}^{2}(\tau)= M_{i}^{2}(1) \tau^{\zeta_i(2)}$. Again, the index $i$ indicates the $i^{th}$ interval over which the exponents are calculated and tracks any (real or statistical) time variation in the value of $\zeta_i(2)$. In an infinitely large interval,  $N\rightarrow\infty$, the limits of the integral $y_i^\pm \rightarrow \pm \infty$; here however each $i^{th}$ interval of the time series will impose different finite extremal values $y_i^\pm$. For the heavy-tailed processes in particular, the statistics of the $y_i^\pm$ can be anticipated to have a significant effect on the statistics of the $\zeta_i(2)$; this has been discussed for the case of $\alpha$- stable L\'evy processes in \cite{Kiyani2006}. These L\'evy processes,  possess heavy tails in the PDFs of their increments, with tails that fall as $P(y)\sim y^{-(1+\alpha)}$ power-laws. The  $\alpha$-stable L\'evy processes have divergent moments for $p=2$ and above;  for a finite sized sample the moments exist but  can be dominated by the behaviour of rare outlying points in the tails which introduce a pathological bias when estimating scaling exponents from the moments \cite{Kiyani2006} (for a wider discussion see \cite{ddw2004}).
We circumvent these difficulties, at least for self-affine time series, by restricting our analysis to the scaling of the second order moment $\zeta(2)$, and by using the iterative conditioning technique \cite{Kiyani2006}. This simple and robust technique for exponent estimation removes a small percentage of the extreme data values which are poorly sampled statistically. In some pathological cases such as the $\alpha$-stable L\'evy distributions these rare extreme points are of the order of and sometimes larger than the whole sum \cite{BardouBouchaud02}. Because they are so large they tend to dominate the statistics and thus the scaling of the higher order moments. This can be clearly seen if we look at the discrete definition of the moments of order $p$
 \begin{equation}
 M_{i}^{p}(\tau)=\frac{1}{N}\sum_{j=1}^{N}\left(y_{i}^{p}\right)_{j}\ . \label{eq:5pt5}
 \end{equation}
The reasoning and full illustration of this iterative conditioning method to heavy-tailed non-Gaussian distributions is discussed in \cite{Kiyani2006}. Although not discussed in this paper, the iterative conditioning technique is also an unbiased robust technique for distinguishing self-affine (monofractal) from multifractal behaviour.

We will focus here on parameter stationarity as opposed to trend stationarity. The former refers to the change in the intrinsic dynamics of the process of interest as characterized by its quantitative statistical properties (the behaviour of the moments); as opposed to the latter which is simply an additive trend to the signal. In particular we will focus on the stationarity of the scaling of the moments as captured by the exponents $\zeta_{i}(p)$. If secular trends are present in the time series then the time series of increments will be approximately trend-free provided our differencing scale $\tau$ is sufficiently small \cite{KantzSchreiber1997}. A secular trend can also be removed by detrending or by the method of studying the scaling of moments of wavelet coefficients where an appropriate wavelet is chosen with a large number of zero-moments \cite{abry1998,abry2003}. The more complex case of mixed dynamics i.e. two or more intrinsically different processes represented in different sections of a time series will not be considered here.

\subsection{Data generation and sources}
We will consider synthetically generated signals that are both stationary and nonstationary with respect to their increments.
The signals with stationary independent increments will consist of a standard Brownian motion and standard symmetric $\alpha$-stable L\'evy motion for four values of the exponent $\alpha$ \cite{samorodnitsky1994,SiegertFriedrich2001}; the latter being highly non-Gaussian and heavy-tailed with very large excursions in their time series. To survey a broad range of such processes we have also included non-Markovian versions of the above processes. These include a long-memory fractional Brownian motion (fBm), and a long-memory persistent and anti-persistent linear fractional stable motion (LFSM) -- see \cite{Beran1994,samorodnitsky1994,stoev2004a,stoev2004b} for more details on these processes and in particular \cite{stoev2004b} for the algorithm and MATLAB code for the LFSM.

We also investigate a multifractal time series generated from a discrete multiplicative cascade  process in the form of the $p$-model \cite{meneveau1987,meneveau1990}. The $p$-model is used as a model for intermittent turbulence \cite{Frisch1995, pagel2002}. The intermittency of the $p$-model time series leads to non-time stationary finite $N$ moments; however the set of scaling exponents $\zeta_{i}(p)$ are stationary.

The nonstationary time series we will consider are a standard Brownian motion with linearly varying standard deviation of its increments with time $\sigma\sim t$, and cyclically varying standard deviation (cyclically stationary) $\sigma\sim sin^{2}(t)$.
All of the above synthetic time series were generated in MATLAB with appropriate random seeding and sample sizes of $N \sim 10^6$.

Lastly, we will consider three real-world time series which have been found to exhibit scaling \cite{freeman2000,hnat2005,hnat2002,Kiyani2007}. These consist of two time series of 100 second resolution magnetic field $B_{z}$ and speed $v$ from the NASA WIND spacecraft at 1AU in the solar minimum year 1996; and a 64 second resolution one year long time series of the magnetic field energy density $B^{2}$ from the NASA ACE spacecraft in the solar maximum year 2000. All of these time series consist of $N\sim 5\times 10^{5}$ data samples and can be downloaded from CDA web \url{http://cdaweb.gsfc.nasa.gov/}.

\section{Results}
\begin{figure}
 \includegraphics[width=0.85\columnwidth,keepaspectratio]{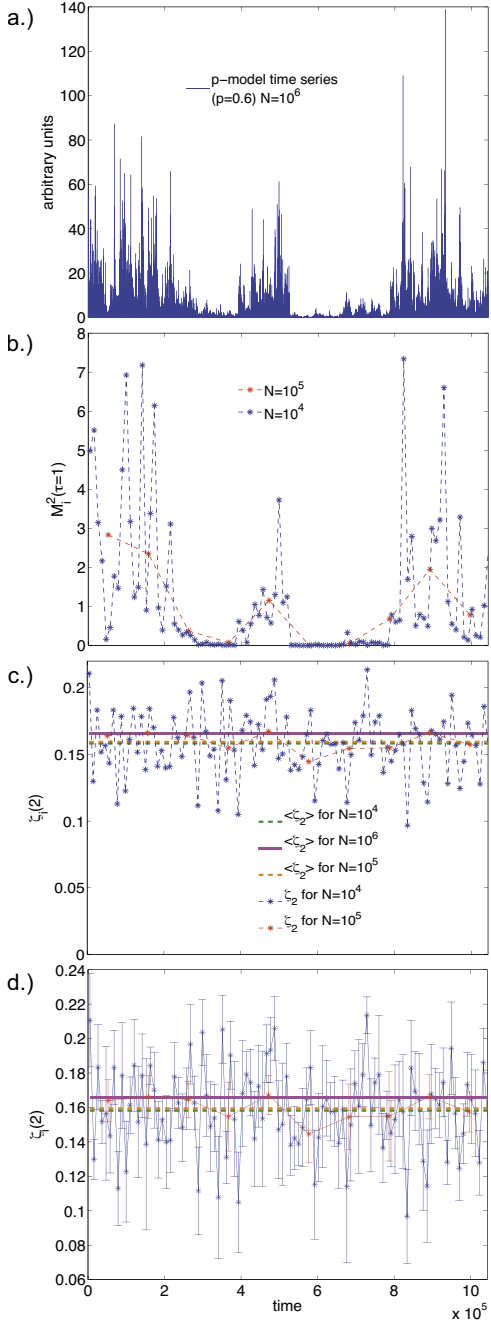}
\caption{\label{fig:1} (Color online) a.) Time series of length $N=10^{6}$ for the $p$-model ($p=0.6$). b.) Variation of the second order moment of the increments $y_{i}(t,\tau)$ for time-scale $\tau=1$ of the above time series where the original time series has been partitioned into $L=100$ and $L=10$ equal sized intervals. c.) Variation of conditioned $\zeta_{i}(2)$ with time for the $p$-model with the same segmentation as in b.) -- also shown are the mean values of the exponents for different partitioning corresponding to different sample sizes; d.) Same as (c.) but with errors explicitly shown.}
\end{figure}

We study the variation of the scaling exponent of the second order moment $\zeta_i(2)$ with sample size $N$.
The process by which the exponent $\zeta_i(2)$ is estimated for $L$ contiguous intervals of $N$ points of a time series is illustrated in Figure \ref{fig:1} for the $p$-model. We begin with the time series in Figure \ref{fig:1}(a) which we parse into $L$ intervals. 
For each of these intervals we obtain an estimate of $\zeta_i(2)$ as the gradient of a
linear least squares regression to a log-log plot of the second-order moment $M_{i}^{2}(\tau)$ versus the scale or differencing parameter $\tau$. This method of obtaining the scaling exponents is also known as the structure function technique \cite{Frisch1995,Hnat2004} and is closely related to variance plot, correlogram and log-periodogram techniques \cite{Beran1994,EmbrechtsMaejima2002} -- in the latter reference \cite{EmbrechtsMaejima2002} it is identical to the variogram technique. We focus on this particular method to estimate  $\zeta_i(2)$  as it provides a point of contact with asymptotic $N \rightarrow \infty$ estimates of the variance of the power spectral exponent $\beta$ which are based on linear regression over a finite range power law power spectrum \cite{Beran1994,robinson1995}. In both cases, the variance in the estimated exponent will depend upon the details of the linear regression. For the second order moment these details include the range of values of $\tau$ over which $M_{i}^{2}(\tau)$ is a power law; the number of different $\tau$ for which we calculate  $M_{i}^{2}(\tau)$ and use in the linear regression; and the uncertainty of each  $M_{i}^{2}(\tau)$ value. 
In all cases considered here we optimize these details to minimize the linear regression error but importantly use the same algorithm for all of the sample time series that we discuss. 

The linear fit is obtained by linear least-squares regression which also provides an estimate of the error. We augment this estimate of the error by varying the start and end points of the regression by a few points and obtaining the difference in the exponents. The linear regression was done over $\sim 20$ values of the scale parameter $\tau$, where $\tau$ was increased geometrically as $\tau=base^k$, where $k \in \lbrace 0,\cdots,40 \rbrace$ and $base$ was chosen to be $1.2$. The fit was done over this reduced set of measurements at $\sim 20$ values of $\tau$ so that a fair comparison can be made with the real-world data (to be discussed later) where only a limited power-law range is seen. 

Due to its highly intermittent nature the $p=0.6$ $p$-model is not time stationary in its finite $N$ moments and this can be seen in Figure \ref{fig:1}(b) where we plot consecutive values of the second order moment $M_{i}^{2}(\tau)$ obtained for each of the $L$ intervals of $N$ points, shown for $\tau=1$ and two values of $N$.
For the $p$-model time series shown here, the second order moment follows the local amplitude of fluctuations in the time series itself; comparing the ratio of the amplitude of these fluctuations to the signal amplitude is one of the classical `first base' techniques for establishing whether the signal is stationary (in the weak sense)\cite{KantzSchreiber1997}. As one would expect from (\ref{eq:5pt5}), this variation of the second-order moment $M_{i}^{2}(\tau)$ with the amplitude of the time series is emphasized as we decrease $N$ as any estimates of the statistics from smaller sample size will naturally mimic the more local features of the time series. This behaviour is more pronounced in very intermittent signals i.e. those with heavy-tailed fluctuation PDFs. 

We also plot in Figures \ref{fig:1}(c) and (d) the corresponding estimates of $\zeta_{i}(2)$ for each interval. These two panels show the same data, that is, the estimates of $\zeta_{i}(2)$ plotted without (c) and with (d) error bars obtained from the linear regression and the error augmentation outlined above.  As intuitively expected, if we decrease the sample size $N$ over which the $\zeta_{i}(2)$ are computed, the scatter increases. However unlike the moments, there is no clear trend with the amplitude of the signal, indicating stationarity of the scaling exponent $\zeta_{i}(2)$. This latter phenomenon will also be encountered in the non-stationary Brownian time series we will study. The estimates of $\zeta_{i}(2)$ can be seen to vary by up to a factor of two for $N=10^4$ for this realization of the $p$-model. This underlies the difficulty of obtaining physically meaningful estimates of scaling exponents for physically realizable $N$.
We can see that the error bars approximately capture the fluctuations in the estimates of $\zeta_{i}(2)$ for the case of the $p$-model. 
As we wish to include strongly non-Gaussian processes in our study, we will henceforth present numerical estimates of the variance of $\zeta_{i}(2)$ obtained directly from computing many values of   $\zeta_{i}(2)$ rather than the linear regression error.

\begin{figure}
 \includegraphics[width=0.99\columnwidth,keepaspectratio]{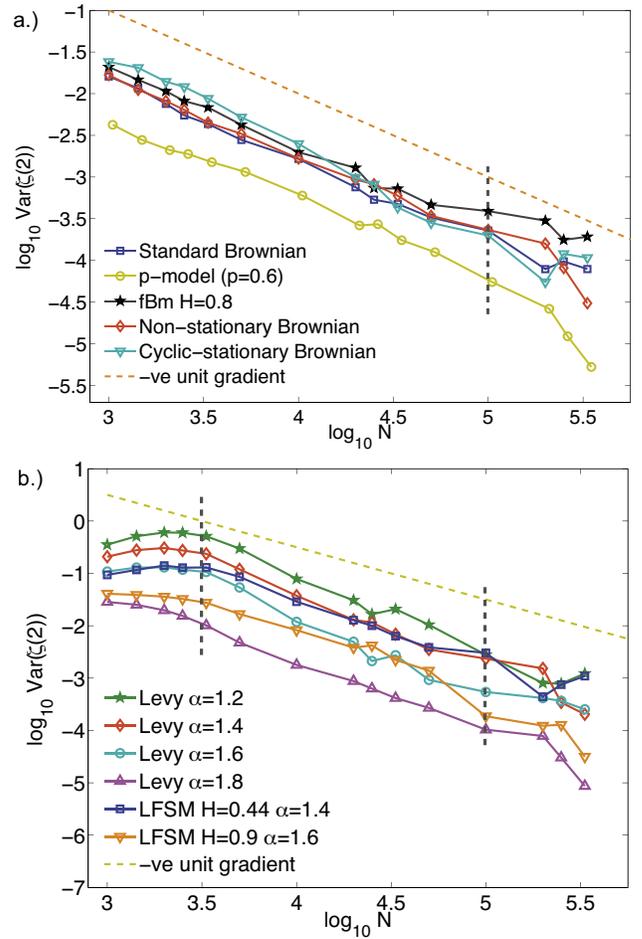}
 \caption{\label{fig:2} (Color online) a.) The variance of conditioned $\zeta(2)$ with sample size $N$ for all the synthetic finite variance processes studied shown on a log-log plot. b.) same as in a.) for all the synthetic infinite variance processes studied. The diagonal dashed line on both these plots indicates a negative slope of unit gradient so that comparison with  theoretically expected asymptotic behaviour can be made. The vertical black dashed lines indicate the areas outside of which errors begin to dominate due to i.) (bottom vertical line) lack of values of $\zeta_i(2)$ to make a decent estimate of $Var(\zeta(2))$; and ii.) (top vertical line) failure of the iterative conditioning technique to obtain unbiased estimates of $\zeta(2)$.}
\end{figure}
The essential point is that quite significant variation in the scaling exponents can arise in time stationary, but intermittent, time series; even when these are estimated over intervals of data that might intuitively be considered to be of adequate length. In order to distinguish variation in the scaling exponents that is statistical (finite $N$ effect) as shown above, from that which reflects intrinsic non-time stationarity, some estimate of the $N$ dependence of the variance of $\zeta(2)$ is needed; this will also point to an estimate of the minimum number of observations $N$ needed to obtain a `reasonably accurate' estimate of  $\zeta(2)$. We will now explore the variance of $\zeta(2)$ as a function of $N$.

In the limit $N\rightarrow\infty$, $\beta$, when estimated via a log-periodogram varies as  $Var[\beta] \sim 1/N$  \cite{Beran1994,robinson1995}. This limiting behaviour is also known for some other estimation schemes of the self-similarity parameter \cite{stoev2002} (as we discuss later here). Thus we would anticipate that for sufficiently large $N$, $Var[\zeta_{i}(2)] \sim 1/N$ for our moment scaling estimation also. However, we do not know the rate of convergence with $N$ to this limiting behaviour and can also anticipate that this will depend upon  whether the PDF of the increments is heavy-tailed, and whether or not the increments are dependent -- both of which introduce further difficulties in obtaining an unbiased estimator.

In Figure \ref{fig:2} we plot the variance of $\zeta_{i}(2)$ against the sample size $N$ on log-log axes, for a range of $N$ that are feasable in realistic realizations of physical systems. Figure \ref{fig:2}(a) shows the behaviour of a subset of our synthetic time series that are intrinsically finite variance processes; Figure \ref{fig:2}(b) shows all the synthetic time series from infinite variance processes that we consider. Plotting these on log-log axes reveals a characteristic power law trend for all the processes:
\begin{equation}
 Var[\zeta(2)]= C N^{-\gamma}\ . \label{eq:8}
\end{equation}
We see that indeed, $\gamma \sim 1$ for the intrinsically finite variance processes.
More pragmatically, we can use this plot to make an estimate of the minimum sample size $N_{min}$ needed in order to estimate $\zeta(2)$ such that the error introduced from the small sample size $N=N_{min}$ is, say, $\sim 5\%$. We propose a simple criterion
\begin{equation}
 \frac{\sqrt{Var[\zeta(2)]}}{\zeta(2)\vert_{L=1}} \lesssim 0.05 \ , \label{eq:6}
\end{equation}
where $\zeta(2)\vert_{L=1}$ is the value of $\zeta(2)$ estimated for the entire time series (assuming that the scaling is stationary). This leads to
\begin{equation}
 Var[\zeta(2)] \lesssim \left( 0.05\zeta(2)\vert_{L=1}\right)^{2}\ , \label{eq:7}
\end{equation}
where the value of $N_{min}$ is extrapolated from the plot of $Var[\zeta(2)]$ Vs. $N$, from Figure \ref{fig:2}(a). For these finite variance processes expressions (\ref{eq:6}) and (\ref{eq:7}) yield $N_{min}\sim 10^{3}$ for the fBm model; $N_{min}\sim 10^{4}$ for the standard Brownian motion (stationary and non-stationary); and $N_{min}\sim 10^{5}$ for the the $p$-model.

 One can invert these relationships to obtain the approximate error on $\zeta_{i}(2)$ given a sample size $N$ from which it was calculated. The constant $C$ in (\ref{eq:8}) is also intrinsic to our estimate of $N_{min}$; operationally the procedure for obtaining the error on $\zeta_{i}(2)$ in this manner would also include estimating $C$ from the plot in Figure \ref{fig:2}(a).

Processes that show scaling often have increments drawn from a heavy tailed PDF, these may also not intrinsically have finite variance as is the case for the $\alpha$-stable L\'evy processes.
 Figure \ref{fig:2}(b) shows the $N$ dependence of all the infinite variance synthetic time series that we have considered, including those with long-range memory. The curves are all generated from time series which possess heavy-tailed PDFs for their increments. These include both the ordinary and fractional L\'evy increments. The curves in Figure \ref{fig:2}(b) have a range of $\gamma$ values close to but also clearly distinct from $\gamma=1$. As will be discussed later this is due to slow convergence to the asymptotic $N^{-1}$ behaviour; from Figure \ref{fig:2}(b) we can see that the L\'evy process which is closest to Gaussian, namely with $\alpha=1.8$, has behaviour closest to $\gamma \sim 1$.
 In a similar way to the method used above for the finite variance synthetic processes, we make  empirical estimates of $N_{min}$ required to obtain an estimate of $\zeta(2)$ to within $\sim5\%$ for the infinite variance processes.
For the L\'evy processes $\alpha=1.2$ and $\alpha=1.4$, and LFSM ($H=0.44$, $\alpha=1.4$) $N_{min}\sim 10^{5}$; for the $\alpha=1.6$ case $N_{min}\sim 10^{4}$; and for the $\alpha=1.8$ case and LFSM ($H=0.9$, $\alpha=1.6$) $N_{min}\sim 10^{3}$. 
The relevant property in the context of estimating the uncertainty on  $\zeta_{i}(2)$ is that for realizable $N$, these processes do not show an $N^{-1}$ dependence. Also, unlike the Gaussian processes in Figure \ref{fig:2}(a) which cluster around a similar $C$ value, the infinite variance processes have noticeably different values of $C$ which depends on both the the tail exponent $\alpha$ and also on the degree of memory in the process given by $H-(1/\alpha)$ \cite{stoev2002}.
 
Finally, combining equations (\ref{eq:8}) and (\ref{eq:7}) we obtain
\begin{equation}
 N_{min} = C^{1/\gamma} \left(0.05\zeta(2)\vert_{L=1}\right) ^{-2/\gamma}\ , \label{eq:9}
\end{equation}
where both $C$ and $\gamma$ depend on the process in question; and for finite variance processes $\gamma=1$. 

\subsubsection*{Error analysis}
\begin{figure*}
 \includegraphics[width=0.8\textwidth,keepaspectratio]{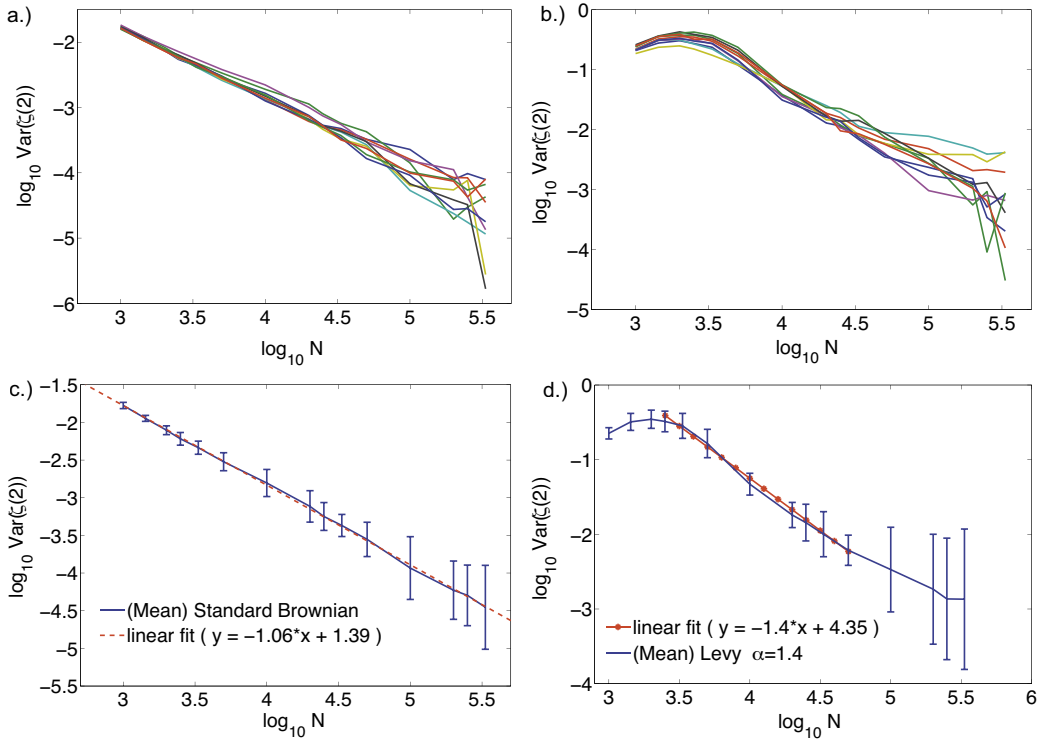}
 \caption{\label{fig:3} (Color online) Plots showing how errors can be ascribed to the plots in Figure \ref{fig:2}. The top plots show the results of the study for 10 different randomly seeded realisations of sample size $N=10^6$ for a.) a standard Brownian motion and b.) an $\alpha$-stable L\'evy motion ($\alpha$=1.4). Plots c.) and d.) are the mean averages of the realisations in a.) and b.) respectively, where the error bars are calculated from the maximum deviation from this mean in the 10 realisations.}
\end{figure*}
To estimate the errors in the estimates of $Var(\zeta(2))$ a small monte-carlo type study was performed in which different random seeds were used to generate 10 different realisations of the two archetypal processes studied here i.e. finite and infinite variance processes in the form of 10 different realisations of a standard Brownian motion and a standard $\alpha$-stable L\'evy process ($\alpha=1.4$). The computation of the $\log Var(\zeta(2))$ Vs. $\log N$ plots were then calculated for each of these realisations; these are shown in Figures \ref{fig:3}(a) and (b). 
We then average over these realisations to obtain an average value of $Var(\zeta(2))$ for each $N$, shown on log-log axes in Figures \ref{fig:3}(c) and (d); the ensemble of realisations also provides an error on this value via the maximum deviation from this average.

At large $N$ errors are dominated by there being fewer values of computed $\zeta_{i}(2)$ and at small $N$, by poor resolution of the PDF from which we ultimately compute $\zeta_{i}(2)$.
In particular, at small $N$ we can see from the plots for the $\alpha$-stable processes that there is a systematic deviation from power law behaviour in $N$. This arises from a breakdown in the iterative conditioning technique \cite{Kiyani2006} at small $N$.

In the next section we will discuss the PDFs of the scaling exponents $\zeta_{i}(2)$ obtained from this study. When these are close to Gaussian, standard Chi-squared distributions and F-test techniques could provide methods of obtaining errors for values of $Var[\zeta(2)]$, even from a single realisation. In this context we should mention the use of bootstrap re-sampling methods for providing distributions, confidence intervals and statistical significance for parameter estimates in situations when one is limited by a single realisation \cite{Wendt2007,Zoubir1998}. Although the convergence and consistent properties of such techniques in the case of heavy-tailed distributions \cite{Hall1990,Athreya1987,Knight1989} and especially infinite-variance processes are still unclear we envisage the use of such methods in future research.

Finally, one could in principle increase the available number of values of $\zeta_{i}(2)$ by overlapping intervals of size $N$ within a given single realisation. We have, however, found that this introduces a significant systematic bias in the computed values of $Var[\zeta(2)]$.

\subsubsection*{Real-world data}
\begin{figure}
 \includegraphics[width=0.99\columnwidth,keepaspectratio]{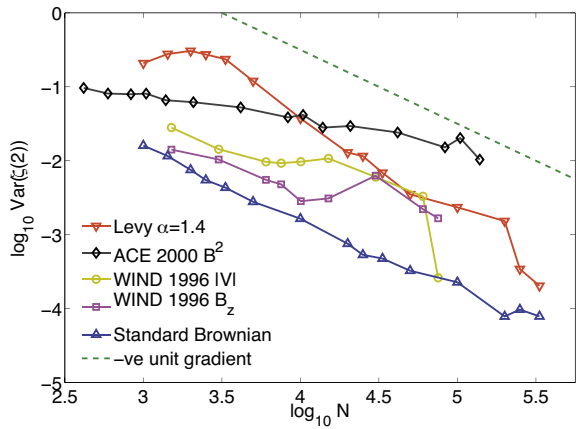}
 \caption{\label{fig:4} (Color online) Plot of the variance of conditioned $\zeta(2)$ with sample size $N$ for all the real-world data sets studied shown on a log-log plot. The dashed line on both of these plots indicates a negative slope of unit gradient. Also included for comparison are the archetypal synthetic data sets for the finite variance and infinite variance processes.}
\end{figure}

We calculate $Var[\zeta(2)]$ values for the examples of real-world data sets discussed earlier in the introduction. The plot detailing this study is shown in Figure \ref{fig:4}. For comparison we have also included on this plot the variation of $Var[\zeta(2)]$ with $N$ for the two archetypal cases of finite and infinite variance processes in the form of a standard Brownian motion and an $\alpha$-stable L\'evy process ($\alpha=1.4$); we also plot a negative unit slope for the asymptotic $N \rightarrow \infty$ behaviour obtained from large sample theory, this is indicated by the dashed line. Figure \ref{fig:4} shows that the real-world data can show significant departures from the synthetic data. 

The WIND data illustrates the effect of large data gaps which are not present in the ACE data shown; this limits the amount of data available for certain $N$ which is reflected in the corresponding estimations of $Var[\zeta(2)]$. For the ACE $B^2$ data we can see a clear systematic departure from the synthetic models. We will discuss this latter data set in the next section below.

The problem with a single length $N$ realisation is that we cannot calculate the errors on $Var[\zeta(2)]$ as done in the previous section; and thus have no way of gauging how close these graphs are to the expected asymptotic behaviour predicted by large-sample theory. However, one can still estimate an error for measurements of $\zeta(2)$ obtained from a finite data size $N$, in the same way as was done in equations (\ref{eq:6})-(\ref{eq:9}). For example, in the case of the ACE $B^2$ data this would indicate that a $N\sim10^5$ sample size would introduce an error of $\sim 12 \% $ in the estimated values of $\zeta(2)$ using the iteratively conditioned moment scaling technique.

\subsection{Underlying statistics of $\zeta(2)$}

\begin{figure*}
\includegraphics[width=0.8\textwidth,keepaspectratio]{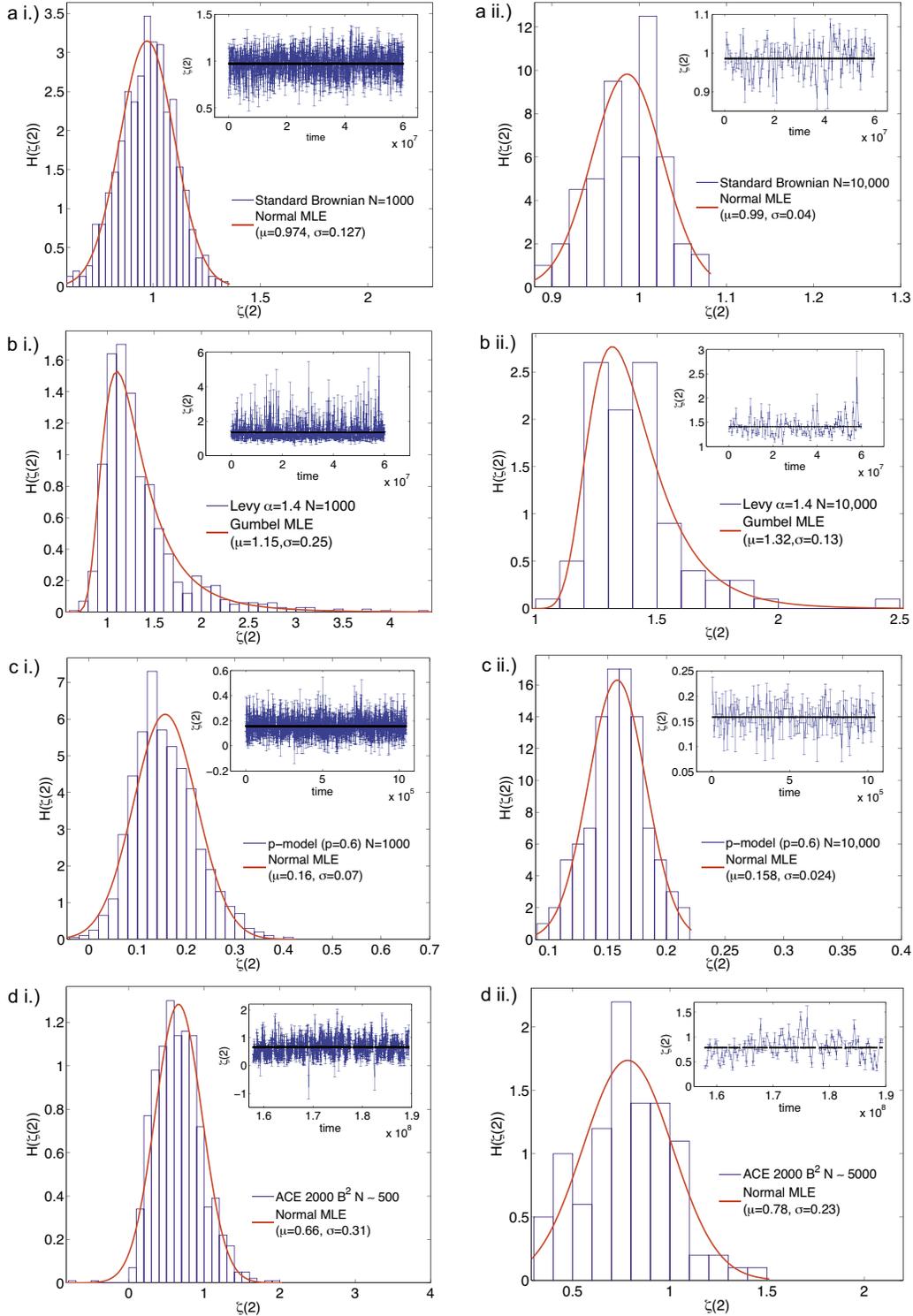}
\caption{\label{fig:5} (Color online) PDF's $H(\zeta(2))$ obtained for (a) a standard Brownian motion, (b) $\alpha$-stable L\'evy process ($\alpha=1.4$), (c) $p$-model (p=0.6) and (d) ACE 2000 $B^2$ for two different sample sizes (i) $N=1000$ ($N=500$ for ACE) and (ii) $N=10,000$ ($N=5000$ for ACE). The sample PDF's are overlaid with Maximum Likelihood Estimate (MLE) model fits of a Normal distribution for a, c and d; and Gumbel max-stable (extreme value type I) fits for b. For both these models $\mu$ is the location parameter and $\sigma$ is the scale parameter, which for the Normal distribution coincide with the mean and standard deviation. The samples of $\zeta_i(2)$ (varying with time $t$) from which these PDFs are constructed, are shown in the corresponding inserts of each plot.}
\end{figure*}

We plot in Figure \ref{fig:5} the PDFs $H\left(\zeta(2)\right)$ for three of the representative models we have studied along with the PDFs $H\left(\zeta(2)\right)$ for one of the real-world data sets. For each of these time series, PDFs have been constructed for two different sample sizes $N$. We see that apart from the $\alpha$-stable case, these PDFs are well described by a Gaussian distribution, as can be seen by the Maximum Likelihood fits. The $\alpha$-stable L\'evy case is shown in Figures \ref{fig:5}b i.) and b ii.) to be well described by a Gumbel max-stable Extreme Value distribution \cite{Gumbel67}. To see why nearly all of our PDFs corresponding to finite variance processes are close to Gaussian we appeal to large sample-theory. 

To facilitate understanding we employ more heuristic arguments at the expense of mathematical rigour. Interested readers can find more on the mathematical details and proofs in  \cite{FoxTaqqu1986,robinson1995} which deal with spectral parameter estimates of strong long-range dependent Gaussian stationary time series; \cite{velasco1999} for a non-stationary generalization of these; \cite{Bardet2000} for a pseudo-variogram estimator (similar to the method in this paper) to long-range dependent Gaussian processes with stationary increments; and the more recent and extensive paper  by Stoev, Pipiras and Taqqu \cite{stoev2002} which extends the proofs and arguments of \cite{Bardet2000} to infinite variance processes in the form of $\alpha$-stable and linear fractional stable processes. This latter reference will be our main source and point of contact for what follows. A survey of many of these papers and parameter estimation techniques can be found in \cite{Beran1994}.

As mentioned above, we have chosen Stoev \emph{et. al.} \cite{stoev2002} as a point of contact from amongst the extensive literature concerning asymptotic large sample behaviour of parameter estimates. This is primarily because this work has dealt with infinite variance processes of the type discussed here; also our moment scaling technique is a particular form of one of the main estimators used in \cite{stoev2002} (see also \cite{Bardet2000}). We have also used Stoev's MATLAB algorithm for generating the LFSM realisations used in this paper. Similar to our study Stoev \emph{et. al.} use a least squares regression on the moments which they refer to as the `power' estimator. However, instead of taking moments of the increments as we do, they take the moments of coefficients for a Finite Impulse Response Transformation (FIRT) of the discrete time-series, which is characterised by a discrete filter of $n$ members. Our increments are one of the simplest forms of these FIRT coefficients if we take the filter to be comprised of a set of $n=2$ members $\left\lbrace -1,1\right\rbrace $. However, any extra benefits of having more than one zero-moment (moments which are equal to zero) will be lost due to this simplicity. This also applies to the wavelet coefficients used in the study of Stoev \emph{et. al.} where our increments result from taking the mother wavelet to be the superposition of two delta functions (one positive, one negative) separated by a scale $\tau$ -- also known in the literature as the `poor man's wavelet' \cite{Vergassola1991}. Also, an important fact to note when comparing the methods of Stoev \emph{et. al.} to ours is that we differ with the `power' method of these authors by using the iterative conditioning technique which by censoring and excluding the poorly sampled large extreme events, correct for the bias which is pathological in the case of heavy-tailed distributions \cite{Kiyani2006}.

Recall that the second order moment is scaling as $M_i^{2}(\tau)=M_i^{2}(1)\tau^{\zeta(2)}$ and we will be estimating $\zeta(2)$ from the gradient of a log-log plot
\begin{equation}
\log M_{i}^{2}(\tau)=\zeta(2)\log\tau+\log M_{i}^{2}(1)\ .\label{eq:10}
\end{equation}

For the discrete data the gradient can be estimated via least squares linear regression, and the problem can be set out as
\begin{equation}
\mathbf{M_{\log}^{2}=T_{\log}Z}+\frac{1}{\sqrt{N}}\mathbf{\epsilon}\ ,\label{eq:11}
\end{equation}
where
\begin{equation}
\mathbf{M_{\log}^{2}}=\left(\begin{array}{c}
\log M_{i}^{2}(\tau_{1})\\
\vdots\\
\log M_{i}^{2}(\tau_{k})\end{array}\right)\ ,
\label{eq:12}
\end{equation}
is the vector of the observations (or dependent variables);
\begin{equation}
\mathbf{T_{\log}}=\left(\begin{array}{cc}
\log\tau_{1} & 1\\
\vdots & \vdots\\
\log\tau_{k} & 1\end{array}\right)\ ,
\label{eq:13}
\end{equation}
contains the vector of the scales (or independent variables);
\begin{equation}
\mathbf{Z}=\left(\begin{array}{c}
\zeta(2)\\
\log M_{i}^{2}(1)\end{array}\right)\ ,
\label{eq:14}
\end{equation}
is the vector of the parameters needed to be estimated; and
\begin{equation}
\mathbf{\epsilon}=\left(\begin{array}{c}
\sqrt{N}\left(\log M_{i}^{2}(\tau_{1})-\log\hat{M}_{i}^{2}(\tau_{1})\right)\\
\vdots\\
\sqrt{N}\left(\log M_{i}^{2}(\tau_{k})-\log\hat{M}_{i}^{2}(\tau_{k})\right)\end{array}\right)\ ,\label{eq:15}
\end{equation}
is the vector of estimation errors between the sample measurements and those of the true expected values (denoted by $\hat{M}$) for which the scaling relation in (\ref{eq:10}) actually holds.
The solution to equation (\ref{eq:11}) is then given by the well known ordinary linear least squares estimator to the parameters as
\begin{equation}
\mathbf{Z=\left( T_{\log}^{t}T_{\log}\right)^{-1} T_{\log}^{t} M_{\log}^{2}} \ ,
\label{eq:15pt5}
\end{equation}
where superscript $t$ represents a matrix transpose. (\ref{eq:15pt5}) is simply a linear combination of the dependent variable $\log M_{i}^{2}(\tau)$ i.e. a sum, so that for $\zeta_{i}(2)$ one can write the ordinary least squares estimate as
\begin{equation}
\zeta_{i}(2)=\sum_{j=1}^{k}a_j\left( \log M_{i}^{2}(\tau_{j}) \right) \ ,
\label{eq:15pt75}
\end{equation}
where the $a_{j}$ are all the elements of the appropriate vector from
$\mathbf{\left( T_{\log}^{t}T_{\log}\right)^{-1} T_{\log}^{t}}$. We will return to this form of the ordinary least squares solution shortly.

Adapted to the notation used in this paper, Theorem 3.1 in \cite{stoev2002} states that if $\zeta(2)$ is the FIRT coefficient estimator (using least squares regression) for the scaling exponent and $\hat{\zeta}(2)$ the true expected value; then as $\lim N\rightarrow \infty$
\begin{equation}
\sqrt{N}\left( \zeta(2)-\hat{\zeta}(2)\right) \rightarrow \mathcal{N}\left( 0,\sigma^{2}\right) 
\label{eq:16}
\end{equation}
where $\mathcal{N}\left( 0,\sigma^{2}\right)$ is a Normal distribution with mean $0$ and variance $\sigma^{2}$. Strictly speaking this theorem requires that the FIRT coefficients obey the following inequality between the number of zero-moments of the FIRT filter, the self-similarity parameter $H$ and the tail (stability) index $\alpha$
\begin{equation}
Q > H + \frac{1}{\alpha \left( \alpha - 1 \right) } \ ,
\label{eq:17}
\end{equation}
which in the case of the ordinary Brownian motion and the $\alpha$-stable L\'evy processes, where $H=1/\alpha$, generalises to
\begin{equation}
Q > \frac{1}{ \left( \alpha - 1 \right) } \ .
\label{eq:18}
\end{equation}
The moment scaling scheme based on the raw increments has only one zero moment, hence $Q=1$; and as a result, except for the $p$-model for which $\alpha$ and $H$ are unknown (or not applicable), the above inequality does not hold for any of the synthetic models. However as mentioned in \cite{Bardet2000} for Gaussian processes, where in equation (\ref{eq:17}) $\alpha=2$, the $Q=1$ case is sufficient for the above theorem to hold as long as $H<0.75$. For our fBm model $H=0.8$ and we find that the results of the theorem (\ref{eq:16}) still hold. Thus we can conjecture that the criterion in (\ref{eq:17}) and (\ref{eq:18}) can be relaxed a little so that the inequality becomes an approximate inequality. Also, Stoev \emph{et. al.} \cite{stoev2002} show via simulations that the estimators continue to work well even when the criterion in (\ref{eq:17}) and (\ref{eq:18})
 are not satisfied. The essential reason why this criteria was initially introduced, was so that the estimator could distinguish between actual long-memory effects and trends \cite{Bardet2000}.

We now consider the exception that we have found to this behaviour -- that of the infinite variance processes at finite $N$. We would expect that in the $N \rightarrow \infty$ limit the results of the above theorem will also hold for the infinite variance processes. However, we believe that in this case the convergence will be slow and will depend upon the number of scales $\tau$ that were used to conduct our linear least squares regression. 

We will now go beyond the above asymptotic arguments to discuss the non-Gaussian intermediate finite $N$ behaviour that we see here in Figure \ref{fig:5} (b-i) and (b-ii).
Recall the expression for $M_{i}^{2}(\tau)$ given in equation (\ref{eq:5pt5}) (for $p=2$). 
We will discuss in detail here the case where the sum in (\ref{eq:5pt5}) consists of \emph{i.i.d.} random variables; this is the case for some of our synthetic time series -- these arguments can be developed for other cases. The PDF of this sum will by the Central Limit Theorem tend to a Gaussian and for finite $N$ will probably take the form of a Pearson $\chi^{2}_{\nu}$ type variable with $\nu$ degrees of freedom (see \cite{Bardet2000} for more details). However, for an infinite variance process, the sum in (\ref{eq:5pt5}) will be dominated by the largest extreme events, which in some cases can be of the order of the rest of the sum \cite{BardouBouchaud02}. This will still be the case even when we have excluded some of these extreme events due to the iterative conditioning scheme. Thus the sum will be distributed as the extreme values of an $\alpha$-stable L\'evy distribution -- which is given by a max-stable Frech\'et distribution (see \cite{Kiyani2006}). Note that this will be the case for any $N$, even in the asymptotic large $N$ case. Without too much detail the form of the PDF $P\left(M^{2}_{i}\right)$ of $M^{2}_{i}$ will be of the type
\begin{equation}
P\left(M^{2}_{i}\right)=\frac{\Lambda}{2\left(M^{2}_{i}\right)^{1+\alpha/2}}
\exp\left(-\frac{\Lambda}{\alpha \left(M^{2}_{i}\right)^{\alpha/2}}\right)\ ,
\label{eq:eqn19}
\end{equation}
where any scale parameters have been absorbed into the $\Lambda$. One can then convert this Frech\'et PDF to a PDF $\tilde{P}\left(M^{2}_{i,\log}\right)$ corresponding to the dependent variable $\log M_{i}^{2}(\tau)$ in the least squares scheme in (\ref{eq:15pt75}), which under a simple conservation of probability will be given by
\begin{equation}
\tilde{P}\left(M^{2}_{i,\log}\right)=\frac{\Lambda}{2}\exp \left(-\frac{\alpha}{2}M^{2}_{i,\log}-\frac{\Lambda}{\alpha} \exp \left(-\frac{\alpha}{2}M^{2}_{i,\log} \right) \right) \ ,
\label{eq:eqn20}
\end{equation}
which is in the form of a Gumbel extreme value distribution; this is another max-stable distribution \cite{Chapman2002}. The Gumbel max-stable PDF has a long slow exponentially decreasing right tail;  this will imply that a sum of random variables such as (\ref{eq:15pt75}), each distributed with this PDF, will eventually tend to a Gaussian distributed random variable, but slowly due to the heavy right tail. This then captures our result in Figures \ref{fig:5} (b-i) and (b-ii), and may also explain why we do not obtain the $\sim N^{-1}$ behaviour in the plots of $\log Var(\zeta(2))$ Vs. $\log N$ for the $\alpha$-stable L\'evy cases.

As discussed above, for the case of the finite variance synthetic time series the $M_{i}^{2}$ will be well described by a Gaussian or (more realistically for finite $N$) a Pearson $\chi_{\nu}^{2}$ PDF. In the same way as was done with the infinite variance processes above, it can be shown that the PDF of $\log M_{i}^{2}$ can be written as a Gumbel min-stable PDF. Gumbel min-stable PDF's have long slow exponentially decreasing left tails, which in our case will be limited by the fact that $\zeta(2)>0$. The right tail of Gumbel min-stable PDFs is more compact with a rapidly decaying exponential tail. Due to the more compact nature of the PDF, a sum of $\log M_{i}^{2}$ variables will tend to a Gaussian under the Central Limit Theorem much faster than the infinite variance processes above, hence explaining why the PDFs $H\left(\zeta(2)\right)$ for the finite variance processes are well described by a Gaussian.

Finally, there is the open question of the behaviour of the real-world data. The ACE $B^2$ data PDFs of $H\left(\zeta(2)\right)$ show that they can be well described by a Gaussian; however the scaling of $Var(\zeta(2))$ with $N$ using our estimation shows a significant deviation from the $N^{-1}$ behaviour implied by (\ref{eq:16}). This will be the topic of future work.

\section{Conclusions}

We have investigated finite-sample size ($N$) effects on quantifying the statistical scaling properties of time series. We focus on the scaling exponent $\zeta(2)$  of the variance or second moment which for a wide class of processes also gives the spectral exponent $\beta$ of the (power law) power spectrum.
If too small a sample size is used then these fluctuations effectively introduce a pseudo-nonstationarity in the estimates for the scaling exponents.
To achieve an error in the exponent of less than $\sim5\%$, we find that the number of data points $N$ needed varies significantly with the details of the underlying process and is in the range of $10^{3}-10^{5}$ for the synthetic models used in this paper.
 The variance in the exponent when computed from an interval of $N$ samples is known to vary as $\sim 1/N$ for $N \rightarrow \infty$; however, the convergence to this behaviour will also depend on the details of the process and more importantly on the parameter estimating technique used. In particular we have shown that heavy tailed processes can be far from this limiting behaviour for observationally realisable $N$.

We have also considered the case where the scaling exponents are time independent, but where there is a secular time dependence in other parameters such as the standard deviation. For the case of a Brownian motion, the estimate of the scaling exponent is not affected by this time dependence. 
It may thus be too premature to reject a time series for scaling analysis simply because of the non-stationarity of certain parameters i.e. a running mean or standard deviation.  This also highlights the need to distinguish time variation in the moments from that in scaling exponents that are derived from them.

We have focussed here on the moment order scaling technique to calculate the scaling exponents in order to highlight the issue of apparent non-time stationarity. Although there exists extensive statistics literature on the asymptotic $N\rightarrow \infty$ limit of various estimation techniques, further work is needed to investigate how these details pass over to the more pressing and pragmatic need for their implications on quantifying scaling in finite and realisable $N$-sized samples.

\begin{acknowledgments}
The authors would like to thank B. Hnat, G. Rowlands, F. M. Poli,
T. Dudok De Wit and V. Keinhorst for helpful discussions and suggestions.
We acknowledge the financial support of the UK STFC and EPSRC; and R. P. Lepping and K. Oglivie for ACE and
WIND data.
\end{acknowledgments}

\bibliographystyle{apsrev}
% \bibliography{StochasticMHD}	

\begin{thebibliography}{49}
\expandafter\ifx\csname natexlab\endcsname\relax\def\natexlab#1{#1}\fi
\expandafter\ifx\csname bibnamefont\endcsname\relax
  \def\bibnamefont#1{#1}\fi
\expandafter\ifx\csname bibfnamefont\endcsname\relax
  \def\bibfnamefont#1{#1}\fi
\expandafter\ifx\csname citenamefont\endcsname\relax
  \def\citenamefont#1{#1}\fi
\expandafter\ifx\csname url\endcsname\relax
  \def\url#1{\texttt{#1}}\fi
\expandafter\ifx\csname urlprefix\endcsname\relax\def\urlprefix{URL }\fi
\providecommand{\bibinfo}[2]{#2}
\providecommand{\eprint}[2][]{\url{#2}}

\bibitem[{\citenamefont{Frisch}(1995)}]{Frisch1995}
\bibinfo{author}{\bibfnamefont{U.}~\bibnamefont{Frisch}},
  \emph{\bibinfo{title}{Turbulence}} (\bibinfo{publisher}{Cambridge University
  Press}, \bibinfo{year}{1995}).

\bibitem[{\citenamefont{Sethna et~al.}(2001)\citenamefont{Sethna, Dahmen, and
  Myers}}]{Sethna2001}
\bibinfo{author}{\bibfnamefont{J.~P.} \bibnamefont{Sethna}},
  \bibinfo{author}{\bibfnamefont{K.~A.} \bibnamefont{Dahmen}},
  \bibnamefont{and} \bibinfo{author}{\bibfnamefont{C.~R.} \bibnamefont{Myers}},
  \bibinfo{journal}{Nature} \textbf{\bibinfo{volume}{410}},
  \bibinfo{pages}{242} (\bibinfo{year}{2001}).

\bibitem[{\citenamefont{Sornette}(2000)}]{Sornette2000}
\bibinfo{author}{\bibfnamefont{D.}~\bibnamefont{Sornette}},
  \emph{\bibinfo{title}{Critical Phenomena in Natural Sciences}}
  (\bibinfo{publisher}{Springer-Verlag}, \bibinfo{year}{2000}).

\bibitem[{\citenamefont{Pope}(2000)}]{Pope2000}
\bibinfo{author}{\bibfnamefont{S.~B.} \bibnamefont{Pope}},
  \emph{\bibinfo{title}{Turbulent Flows}} (\bibinfo{publisher}{Cambridge
  University Press}, \bibinfo{year}{2000}).

\bibitem[{\citenamefont{Hnat et~al.}(2004)\citenamefont{Hnat, Chapman, and
  Rowlands}}]{Hnat2004}
\bibinfo{author}{\bibfnamefont{B.}~\bibnamefont{Hnat}},
  \bibinfo{author}{\bibfnamefont{S.~C.} \bibnamefont{Chapman}},
  \bibnamefont{and} \bibinfo{author}{\bibfnamefont{G.}~\bibnamefont{Rowlands}},
  \bibinfo{journal}{Physics of Plasmas} \textbf{\bibinfo{volume}{11}},
  \bibinfo{pages}{1326} (\bibinfo{year}{2004}).

\bibitem[{\citenamefont{Smith et~al.}(2006)\citenamefont{Smith, Hamilton,
  Vasquez, and Leamon}}]{Smith2006}
\bibinfo{author}{\bibfnamefont{C.~W.} \bibnamefont{Smith}},
  \bibinfo{author}{\bibfnamefont{K.}~\bibnamefont{Hamilton}},
  \bibinfo{author}{\bibfnamefont{B.~J.} \bibnamefont{Vasquez}},
  \bibnamefont{and} \bibinfo{author}{\bibfnamefont{R.~J.}
  \bibnamefont{Leamon}}, \bibinfo{journal}{Astrophys. J.}
  \textbf{\bibinfo{volume}{645}}, \bibinfo{pages}{L85} (\bibinfo{year}{2006}).

\bibitem[{\citenamefont{Retin\`o et~al.}(2007)\citenamefont{Retin\`o,
  Sundkvist, Vaivads, Mozer, Andr\'e, and Owen}}]{Retino2007}
\bibinfo{author}{\bibfnamefont{A.}~\bibnamefont{Retin\`o}},
  \bibinfo{author}{\bibfnamefont{D.}~\bibnamefont{Sundkvist}},
  \bibinfo{author}{\bibfnamefont{A.}~\bibnamefont{Vaivads}},
  \bibinfo{author}{\bibfnamefont{F.}~\bibnamefont{Mozer}},
  \bibinfo{author}{\bibfnamefont{M.}~\bibnamefont{Andr\'e}}, \bibnamefont{and}
  \bibinfo{author}{\bibfnamefont{C.~J.} \bibnamefont{Owen}},
  \bibinfo{journal}{Nature Physics} \textbf{\bibinfo{volume}{3}},
  \bibinfo{pages}{236} (\bibinfo{year}{2007}).

\bibitem[{\citenamefont{Sundkvist et~al.}(2007)\citenamefont{Sundkvist,
  Retin\`o, Vaivads, and Bale}}]{Sundkvist2007}
\bibinfo{author}{\bibfnamefont{D.}~\bibnamefont{Sundkvist}},
  \bibinfo{author}{\bibfnamefont{A.}~\bibnamefont{Retin\`o}},
  \bibinfo{author}{\bibfnamefont{A.}~\bibnamefont{Vaivads}}, \bibnamefont{and}
  \bibinfo{author}{\bibfnamefont{S.~D.} \bibnamefont{Bale}},
  \bibinfo{journal}{Phys. Rev. Lett.} \textbf{\bibinfo{volume}{99}},
  \bibinfo{pages}{025004} (\bibinfo{year}{2007}).

\bibitem[{\citenamefont{Sorriso-Valvo et~al.}(2007)\citenamefont{Sorriso-Valvo,
  Marino, Carbone, Noullez, Lepreti, Veltri, Bruno, Bavassano, and
  Pietropaolo}}]{Sorriso-valvo2007}
\bibinfo{author}{\bibfnamefont{L.}~\bibnamefont{Sorriso-Valvo}},
  \bibinfo{author}{\bibfnamefont{R.}~\bibnamefont{Marino}},
  \bibinfo{author}{\bibfnamefont{V.}~\bibnamefont{Carbone}},
  \bibinfo{author}{\bibfnamefont{A.}~\bibnamefont{Noullez}},
  \bibinfo{author}{\bibfnamefont{F.}~\bibnamefont{Lepreti}},
  \bibinfo{author}{\bibfnamefont{P.}~\bibnamefont{Veltri}},
  \bibinfo{author}{\bibfnamefont{R.}~\bibnamefont{Bruno}},
  \bibinfo{author}{\bibfnamefont{B.}~\bibnamefont{Bavassano}},
  \bibnamefont{and}
  \bibinfo{author}{\bibfnamefont{E.}~\bibnamefont{Pietropaolo}},
  \bibinfo{journal}{Phys. Rev. Lett.} \textbf{\bibinfo{volume}{99}},
  \bibinfo{pages}{115001} (\bibinfo{year}{2007}).

\bibitem[{\citenamefont{Nicol et~al.}(2008)\citenamefont{Nicol, Chapman, and
  Dendy}}]{nicol2008}
\bibinfo{author}{\bibfnamefont{R.~M.} \bibnamefont{Nicol}},
  \bibinfo{author}{\bibfnamefont{S.~C.} \bibnamefont{Chapman}},
  \bibnamefont{and} \bibinfo{author}{\bibfnamefont{R.~O.} \bibnamefont{Dendy}},
  \bibinfo{journal}{Astrophys. J.} \textbf{\bibinfo{volume}{679}},
  \bibinfo{pages}{862} (\bibinfo{year}{2008}).

\bibitem[{\citenamefont{Jankovicova et~al.}(2008)\citenamefont{Jankovicova,
  Voros, and Simkanin}}]{Jankovicova2008}
\bibinfo{author}{\bibfnamefont{D.}~\bibnamefont{Jankovicova}},
  \bibinfo{author}{\bibfnamefont{Z.}~\bibnamefont{Voros}}, \bibnamefont{and}
  \bibinfo{author}{\bibfnamefont{J.}~\bibnamefont{Simkanin}},
  \bibinfo{journal}{Nonlin. Proc. Geophys.} \textbf{\bibinfo{volume}{15}},
  \bibinfo{pages}{53} (\bibinfo{year}{2008}).

\bibitem[{\citenamefont{Abel and Freeman}(2002)}]{abel2002}
\bibinfo{author}{\bibfnamefont{G.~A.} \bibnamefont{Abel}} \bibnamefont{and}
  \bibinfo{author}{\bibfnamefont{M.~P.} \bibnamefont{Freeman}},
  \bibinfo{journal}{J. Geophys. Res.} \textbf{\bibinfo{volume}{107}},
  \bibinfo{pages}{1470} (\bibinfo{year}{2002}).

\bibitem[{\citenamefont{Mallat}(1999)}]{Mallat1999}
\bibinfo{author}{\bibfnamefont{S.}~\bibnamefont{Mallat}},
  \emph{\bibinfo{title}{A wavelet tour of signal processing}}
  (\bibinfo{publisher}{Academic Press Inc.}, \bibinfo{year}{1999}).

\bibitem[{\citenamefont{Popivanov and Mineva}(1999)}]{Popivanov1999}
\bibinfo{author}{\bibfnamefont{D.}~\bibnamefont{Popivanov}} \bibnamefont{and}
  \bibinfo{author}{\bibfnamefont{A.}~\bibnamefont{Mineva}},
  \bibinfo{journal}{Mathematical Biosciences} \textbf{\bibinfo{volume}{157}},
  \bibinfo{pages}{303} (\bibinfo{year}{1999}).

\bibitem[{\citenamefont{Leontitsis and Vorlow}(2006)}]{Leontitsis2006}
\bibinfo{author}{\bibfnamefont{A.}~\bibnamefont{Leontitsis}} \bibnamefont{and}
  \bibinfo{author}{\bibfnamefont{C.~E.} \bibnamefont{Vorlow}},
  \bibinfo{journal}{Physica A} \textbf{\bibinfo{volume}{368}},
  \bibinfo{pages}{522} (\bibinfo{year}{2006}).

\bibitem[{\citenamefont{Robinson}(1995)}]{robinson1995}
\bibinfo{author}{\bibfnamefont{P.~M.} \bibnamefont{Robinson}},
  \bibinfo{journal}{Ann. Stat.} \textbf{\bibinfo{volume}{23}},
  \bibinfo{pages}{1048} (\bibinfo{year}{1995}).

\bibitem[{\citenamefont{Beran}(1994)}]{Beran1994}
\bibinfo{author}{\bibfnamefont{J.}~\bibnamefont{Beran}},
  \emph{\bibinfo{title}{Statistics for Long-Memory Processes}}
  (\bibinfo{publisher}{Chapman {\&} Hall}, \bibinfo{year}{1994}).

\bibitem[{\citenamefont{Muzy et~al.}(1993)\citenamefont{Muzy, Bacry, and
  Arneodo}}]{Muzy1993}
\bibinfo{author}{\bibfnamefont{J.~F.} \bibnamefont{Muzy}},
  \bibinfo{author}{\bibfnamefont{E.}~\bibnamefont{Bacry}}, \bibnamefont{and}
  \bibinfo{author}{\bibfnamefont{A.}~\bibnamefont{Arneodo}},
  \bibinfo{journal}{Phys. Rev. E} \textbf{\bibinfo{volume}{47}},
  \bibinfo{pages}{875} (\bibinfo{year}{1993}).

\bibitem[{\citenamefont{Abry et~al.}(2003)\citenamefont{Abry, Flandrin, Taqqu,
  and Veitch}}]{abry2003}
\bibinfo{author}{\bibfnamefont{P.}~\bibnamefont{Abry}},
  \bibinfo{author}{\bibfnamefont{P.}~\bibnamefont{Flandrin}},
  \bibinfo{author}{\bibfnamefont{M.}~\bibnamefont{Taqqu}}, \bibnamefont{and}
  \bibinfo{author}{\bibfnamefont{D.}~\bibnamefont{Veitch}},
  \emph{\bibinfo{title}{Theory and applications of long-range dependence}} 
  (\bibinfo{publisher}{Birkhauser}, \bibinfo{year}{2003}), p.
  \bibinfo{pages}{527}.

\bibitem[{\citenamefont{Stoev et~al.}(2002)\citenamefont{Stoev, Pipiras, and
  Taqqu}}]{stoev2002}
\bibinfo{author}{\bibfnamefont{S.}~\bibnamefont{Stoev}},
  \bibinfo{author}{\bibfnamefont{V.}~\bibnamefont{Pipiras}}, \bibnamefont{and}
  \bibinfo{author}{\bibfnamefont{M.~S.} \bibnamefont{Taqqu}},
  \bibinfo{journal}{Signal Proc.} \textbf{\bibinfo{volume}{82}},
  \bibinfo{pages}{1873} (\bibinfo{year}{2002}).

\bibitem[{\citenamefont{Bardet}(2000)}]{Bardet2000}
\bibinfo{author}{\bibfnamefont{J.-M.} \bibnamefont{Bardet}},
  \bibinfo{journal}{J. Time Series Analysis} \textbf{\bibinfo{volume}{21}},
  \bibinfo{pages}{497} (\bibinfo{year}{2000}).

\bibitem[{\citenamefont{Samorodnitsky and Taqqu}(1994)}]{samorodnitsky1994}
\bibinfo{author}{\bibfnamefont{G.}~\bibnamefont{Samorodnitsky}}
  \bibnamefont{and} \bibinfo{author}{\bibfnamefont{M.~S.} \bibnamefont{Taqqu}},
  \emph{\bibinfo{title}{Stable non-Gaussian random processes}}
  (\bibinfo{publisher}{Chapman {\&} Hall}, \bibinfo{year}{1994}).

\bibitem[{\citenamefont{Stoev and Taqqu}(2004{\natexlab{a}})}]{stoev2004a}
\bibinfo{author}{\bibfnamefont{S.}~\bibnamefont{Stoev}} \bibnamefont{and}
  \bibinfo{author}{\bibfnamefont{M.~S.} \bibnamefont{Taqqu}},
  \bibinfo{journal}{Adv. Appl. Prob.} \textbf{\bibinfo{volume}{36}},
  \bibinfo{pages}{1085} (\bibinfo{year}{2004}{\natexlab{a}}).

\bibitem[{\citenamefont{Stoev and Taqqu}(2004{\natexlab{b}})}]{stoev2004b}
\bibinfo{author}{\bibfnamefont{S.}~\bibnamefont{Stoev}} \bibnamefont{and}
  \bibinfo{author}{\bibfnamefont{M.~S.} \bibnamefont{Taqqu}},
  \bibinfo{journal}{Fractals} \textbf{\bibinfo{volume}{12}},
  \bibinfo{pages}{95} (\bibinfo{year}{2004}{\natexlab{b}}).

\bibitem[{\citenamefont{Meneveau and Sreenivasan}(1987)}]{meneveau1987}
\bibinfo{author}{\bibfnamefont{C.}~\bibnamefont{Meneveau}} \bibnamefont{and}
  \bibinfo{author}{\bibfnamefont{K.~R.} \bibnamefont{Sreenivasan}},
  \bibinfo{journal}{Phys. Rev. Lett.} \textbf{\bibinfo{volume}{59}},
  \bibinfo{pages}{1424} (\bibinfo{year}{1987}).

\bibitem[{\citenamefont{Meneveau et~al.}(1990)\citenamefont{Meneveau,
  Sreenivasan, Kailasnath, and Fan}}]{meneveau1990}
\bibinfo{author}{\bibfnamefont{C.}~\bibnamefont{Meneveau}},
  \bibinfo{author}{\bibfnamefont{K.~R.} \bibnamefont{Sreenivasan}},
  \bibinfo{author}{\bibfnamefont{P.}~\bibnamefont{Kailasnath}},
  \bibnamefont{and} \bibinfo{author}{\bibfnamefont{M.~S.} \bibnamefont{Fan}},
  \bibinfo{journal}{Phys. Rev. A} \textbf{\bibinfo{volume}{41}},
  \bibinfo{pages}{894} (\bibinfo{year}{1990}).

\bibitem[{\citenamefont{Kiyani et~al.}(2006)\citenamefont{Kiyani, Chapman, and
  Hnat}}]{Kiyani2006}
\bibinfo{author}{\bibfnamefont{K.}~\bibnamefont{Kiyani}},
  \bibinfo{author}{\bibfnamefont{S.~C.} \bibnamefont{Chapman}},
  \bibnamefont{and} \bibinfo{author}{\bibfnamefont{B.}~\bibnamefont{Hnat}},
  \bibinfo{journal}{Phys. Rev. E} \textbf{\bibinfo{volume}{74}},
  \bibinfo{pages}{051122} (\bibinfo{year}{2006}).

\bibitem[{\citenamefont{Weron}(2002)}]{Weron2002}
\bibinfo{author}{\bibfnamefont{R.}~\bibnamefont{Weron}},
  \bibinfo{journal}{Physica A} \textbf{\bibinfo{volume}{312}},
  \bibinfo{pages}{285} (\bibinfo{year}{2002}).

\bibitem[{\citenamefont{Velasco}(1999)}]{velasco1999}
\bibinfo{author}{\bibfnamefont{C.}~\bibnamefont{Velasco}}, \bibinfo{journal}{J.
  Econometrics} \textbf{\bibinfo{volume}{91}}, \bibinfo{pages}{325}
  (\bibinfo{year}{1999}).

\bibitem[{\citenamefont{Embrechts and Maejima}(2002)}]{EmbrechtsMaejima2002}
\bibinfo{author}{\bibfnamefont{P.}~\bibnamefont{Embrechts}} \bibnamefont{and}
  \bibinfo{author}{\bibfnamefont{M.}~\bibnamefont{Maejima}},
  \emph{\bibinfo{title}{Selfsimilar Processes}} (\bibinfo{publisher}{Princeton
  University Press}, \bibinfo{year}{2002}).

\bibitem[{\citenamefont{{Dudok De Wit}}(2004)}]{ddw2004}
\bibinfo{author}{\bibfnamefont{T.}~\bibnamefont{{Dudok De Wit}}},
  \bibinfo{journal}{Phys. Rev. E} \textbf{\bibinfo{volume}{70}},
  \bibinfo{pages}{055302(R)} (\bibinfo{year}{2004}).

\bibitem[{\citenamefont{Bardou et~al.}(2002)\citenamefont{Bardou, Bouchaud,
  Aspect, and Cohen-Tannoudji}}]{BardouBouchaud02}
\bibinfo{author}{\bibfnamefont{F.}~\bibnamefont{Bardou}},
  \bibinfo{author}{\bibfnamefont{J.}~\bibnamefont{Bouchaud}},
  \bibinfo{author}{\bibfnamefont{A.}~\bibnamefont{Aspect}}, \bibnamefont{and}
  \bibinfo{author}{\bibfnamefont{C.}~\bibnamefont{Cohen-Tannoudji}},
  \emph{\bibinfo{title}{L{\'e}vy Statistics and Laser Cooling}}
  (\bibinfo{publisher}{Cambridge University Press}, \bibinfo{year}{2002}).

\bibitem[{\citenamefont{Kantz and Schreiber}(1997)}]{KantzSchreiber1997}
\bibinfo{author}{\bibfnamefont{H.}~\bibnamefont{Kantz}} \bibnamefont{and}
  \bibinfo{author}{\bibfnamefont{T.}~\bibnamefont{Schreiber}},
  \emph{\bibinfo{title}{Nonlinear Time Series Analysis}}
  (\bibinfo{publisher}{Cambridge University Press}, \bibinfo{year}{1997}).

\bibitem[{\citenamefont{Abry and Veitch}(1998)}]{abry1998}
\bibinfo{author}{\bibfnamefont{P.}~\bibnamefont{Abry}} \bibnamefont{and}
  \bibinfo{author}{\bibfnamefont{D.}~\bibnamefont{Veitch}},
  \bibinfo{journal}{IEEE Trans. Inf. Th.} \textbf{\bibinfo{volume}{44}},
  \bibinfo{pages}{2} (\bibinfo{year}{1998}).

\bibitem[{\citenamefont{Siegert and Friedrich}(2001)}]{SiegertFriedrich2001}
\bibinfo{author}{\bibfnamefont{S.}~\bibnamefont{Siegert}} \bibnamefont{and}
  \bibinfo{author}{\bibfnamefont{R.}~\bibnamefont{Friedrich}},
  \bibinfo{journal}{Phys. Rev. E} \textbf{\bibinfo{volume}{64}}
  (\bibinfo{year}{2001}).

\bibitem[{\citenamefont{Pagel and Balogh}(2002)}]{pagel2002}
\bibinfo{author}{\bibfnamefont{C.}~\bibnamefont{Pagel}} \bibnamefont{and}
  \bibinfo{author}{\bibfnamefont{A.}~\bibnamefont{Balogh}},
  \bibinfo{journal}{J. Geophys. Res.} \textbf{\bibinfo{volume}{107}},
  \bibinfo{pages}{1178} (\bibinfo{year}{2002}).

\bibitem[{\citenamefont{Freeman et~al.}(2000)\citenamefont{Freeman, Watkins,
  and Riley}}]{freeman2000}
\bibinfo{author}{\bibfnamefont{M.~P.} \bibnamefont{Freeman}},
  \bibinfo{author}{\bibfnamefont{N.~W.} \bibnamefont{Watkins}},
  \bibnamefont{and} \bibinfo{author}{\bibfnamefont{D.~J.} \bibnamefont{Riley}},
  \bibinfo{journal}{Phys. Rev. E} \textbf{\bibinfo{volume}{62}},
  \bibinfo{pages}{8794} (\bibinfo{year}{2000}).

\bibitem[{\citenamefont{Hnat et~al.}(2005)\citenamefont{Hnat, Chapman, and
  Rowlands}}]{hnat2005}
\bibinfo{author}{\bibfnamefont{B.}~\bibnamefont{Hnat}},
  \bibinfo{author}{\bibfnamefont{S.~C.} \bibnamefont{Chapman}},
  \bibnamefont{and} \bibinfo{author}{\bibfnamefont{G.}~\bibnamefont{Rowlands}},
  \bibinfo{journal}{J. Geophys. Res.} \textbf{\bibinfo{volume}{110}}
  (\bibinfo{year}{2005}).

\bibitem[{\citenamefont{Hnat et~al.}(2002)\citenamefont{Hnat, Chapman,
  Rowlands, Watkins, and Farrell}}]{hnat2002}
\bibinfo{author}{\bibfnamefont{B.}~\bibnamefont{Hnat}},
  \bibinfo{author}{\bibfnamefont{S.~C.} \bibnamefont{Chapman}},
  \bibinfo{author}{\bibfnamefont{G.}~\bibnamefont{Rowlands}},
  \bibinfo{author}{\bibfnamefont{N.~W.} \bibnamefont{Watkins}},
  \bibnamefont{and} \bibinfo{author}{\bibfnamefont{W.~M.}
  \bibnamefont{Farrell}}, \bibinfo{journal}{Geophys. Res. Lett.}
  \textbf{\bibinfo{volume}{29}} (\bibinfo{year}{2002}).

\bibitem[{\citenamefont{Kiyani et~al.}(2007)\citenamefont{Kiyani, Chapman,
  Hnat, and Nicol}}]{Kiyani2007}
\bibinfo{author}{\bibfnamefont{K.}~\bibnamefont{Kiyani}},
  \bibinfo{author}{\bibfnamefont{S.~C.} \bibnamefont{Chapman}},
  \bibinfo{author}{\bibfnamefont{B.}~\bibnamefont{Hnat}}, \bibnamefont{and}
  \bibinfo{author}{\bibfnamefont{R.~M.} \bibnamefont{Nicol}},
  \bibinfo{journal}{Phys. Rev. Lett.} \textbf{\bibinfo{volume}{98}},
  \bibinfo{pages}{211101} (\bibinfo{year}{2007}).

\bibitem[{\citenamefont{Wendt and Abry}(2007)}]{Wendt2007}
\bibinfo{author}{\bibfnamefont{H.}~\bibnamefont{Wendt}} \bibnamefont{and}
  \bibinfo{author}{\bibfnamefont{P.}~\bibnamefont{Abry}},
  \bibinfo{journal}{IEEE Trans. Sig. Proc.} \textbf{\bibinfo{volume}{55}},
  \bibinfo{pages}{4811} (\bibinfo{year}{2007}).

\bibitem[{\citenamefont{Zoubir and Boashash}(1998)}]{Zoubir1998}
\bibinfo{author}{\bibfnamefont{A.~M.} \bibnamefont{Zoubir}} \bibnamefont{and}
  \bibinfo{author}{\bibfnamefont{B.}~\bibnamefont{Boashash}},
  \bibinfo{journal}{IEEE Sig. Proc. Mag.} \textbf{\bibinfo{volume}{15}},
  \bibinfo{pages}{56} (\bibinfo{year}{1998}).

\bibitem[{\citenamefont{Hall}(1990)}]{Hall1990}
\bibinfo{author}{\bibfnamefont{P.}~\bibnamefont{Hall}}, \bibinfo{journal}{Ann.
  Prob.} \textbf{\bibinfo{volume}{18}}, \bibinfo{pages}{1342}
  (\bibinfo{year}{1990}).

\bibitem[{\citenamefont{Athreya}(1987)}]{Athreya1987}
\bibinfo{author}{\bibfnamefont{K.~B.} \bibnamefont{Athreya}},
  \bibinfo{journal}{Ann. Stat.} \textbf{\bibinfo{volume}{15}},
  \bibinfo{pages}{724} (\bibinfo{year}{1987}).

\bibitem[{\citenamefont{Knight}(1989)}]{Knight1989}
\bibinfo{author}{\bibfnamefont{K.}~\bibnamefont{Knight}},
  \bibinfo{journal}{Ann. Stat.} \textbf{\bibinfo{volume}{17}},
  \bibinfo{pages}{1168} (\bibinfo{year}{1989}).

\bibitem[{\citenamefont{Gumbel}(1967)}]{Gumbel67}
\bibinfo{author}{\bibfnamefont{E.~J.} \bibnamefont{Gumbel}},
  \emph{\bibinfo{title}{Statistics of Extremes}} (\bibinfo{publisher}{Columbia
  University Press}, \bibinfo{year}{1967}).

\bibitem[{\citenamefont{Fox and Taqqu}(1986)}]{FoxTaqqu1986}
\bibinfo{author}{\bibfnamefont{R.}~\bibnamefont{Fox}} \bibnamefont{and}
  \bibinfo{author}{\bibfnamefont{M.~S.} \bibnamefont{Taqqu}},
  \bibinfo{journal}{Ann. Stat.} \textbf{\bibinfo{volume}{14}},
  \bibinfo{pages}{517} (\bibinfo{year}{1986}).

\bibitem[{\citenamefont{Vergassola and Frisch}(1991)}]{Vergassola1991}
\bibinfo{author}{\bibfnamefont{M.}~\bibnamefont{Vergassola}} \bibnamefont{and}
  \bibinfo{author}{\bibfnamefont{U.}~\bibnamefont{Frisch}},
  \bibinfo{journal}{Physica D} \textbf{\bibinfo{volume}{54}},
  \bibinfo{pages}{58} (\bibinfo{year}{1991}).

\bibitem[{\citenamefont{Chapman et~al.}(2002)\citenamefont{Chapman, Rowlands,
  and Watkins}}]{Chapman2002}
\bibinfo{author}{\bibfnamefont{S.~C.} \bibnamefont{Chapman}},
  \bibinfo{author}{\bibfnamefont{G.}~\bibnamefont{Rowlands}}, \bibnamefont{and}
  \bibinfo{author}{\bibfnamefont{N.~W.} \bibnamefont{Watkins}},
  \bibinfo{journal}{Nonlin. Proc. Geophys.} \textbf{\bibinfo{volume}{9}},
  \bibinfo{pages}{409} (\bibinfo{year}{2002}).

\end{thebibliography}

\end{document}